\documentclass[letter]{aa} 
\usepackage[varg]{txfonts}
\usepackage{hyperref}
\usepackage{natbib}
\usepackage{graphicx}

\begin{document} 

   \title{Evidence for anisotropy of cosmic acceleration}
   \author{Jacques Colin\inst{1}
          \and
          Roya Mohayaee\inst{1}
          \and
          Mohamed Rameez\inst{2}
          \and
          Subir Sarkar\inst{3}
          }
   \institute{CNRS, UPMC, Institut d'Astrophysique de Paris
              98 bis Bld Arago, Paris, France\\
         \and
             Niels Bohr Institute, University of Copenhagen, 
             Blegdamsvej 17, 2100 Copenhagen, Denmark\\
         \and
          	Rudolf Peierls Centre for Theoretical Physics, University of Oxford, 
          	Parks Road, Oxford OX1 3PU, United Kingdom\\
             }
   \date{}

  \abstract
    {Observations reveal a `bulk flow' in the local Universe which is faster and extends to much larger scales than is expected around a typical observer in the standard $\Lambda$CDM cosmology. This is expected to result in a scale-dependent dipolar modulation of the acceleration of the expansion rate inferred from observations of objects within the bulk flow. From a maximum-likelihood analysis of the Joint Lightcurve Analysis (JLA) catalogue of Type~Ia supernovae we find that the deceleration parameter, in addition to a small monopole, indeed has a much bigger dipole component aligned with the CMB dipole which falls exponentially with redshift $z$: $q_0 = q_\mathrm{m} + \vec{q}_\mathrm{d}.\hat{n}\exp(-z/S)$. The best fit to data yields $q_\mathrm{d} = -8.03$ and $S = 0.0262~(\Rightarrow d\sim100~\mathrm{Mpc})$, rejecting isotropy ($q_\mathrm{d}=0$) with $3.9\sigma$ statistical significance, while $q_\mathrm{m}=-0.157$ and consistent with no acceleration ($q_\mathrm{m}=0$) at $1.4\sigma$. Thus the cosmic acceleration deduced from supernovae may be an artefact of our being non-Copernican observers, rather than evidence for a dominant component of `dark energy' in the Universe.}
   \maketitle

\section{Introduction}

The foundations of the current standard model of cosmology date back nearly a century to when there was essentially no data available. In particular the Universe was assumed to be exactly isotropic and homogeneous, with space-time described by the maximally symmetric Friedmann--Lema\^{\i}tre--Robertson--Walker metric, and occupied by ideal fluids with purely diagonal energy-momentum tensors~\citep{Peebles:1994xt}. Subsequently it has been recognised that the distribution of galaxies, which is a biased tracer of the underlying distribution of the dominant dark matter, is in fact rather inhomogeneous. Counts-in-spheres of galaxy catalogues suggest that there is a transition to (statistical) homogeneity on scales exceeding $\sim 100$~Mpc \citep{Hogg:2004vw,Scrimgeour:2012wt} although sufficiently large volumes have not yet been surveyed to establish this definitively. This is however the expectation in the current standard cosmological model if the observed large-scale structure has grown under gravity in the sea of dark matter, starting with an initially gaussian random field of small density perturbations with an approximately scale-invariant spectrum. Detailed observations of the temperature fluctuations in the cosmic microwave background (CMB) have broadly confirmed this model~\citep{Akrami:2018vks}. However several anomalies have been noted, e.g. the lack of correlations on large angular scales, the quadrupole-octupole alignment, and the hemispherical power asymmetry --- which seem to imply a violation of statistical isotropy and scale-invariance of primordial  perturbations --- although there is no consensus yet on either their physical nature or their origin \citep{Schwarz:2015cma}.

In our real Universe there are `peculiar motions' due to the local inhomogeneity and anisotropy of surrounding structure. These are non-negligible, e.g. our Local Group of galaxies moves with respect to the universal expansion at $620 \pm 15 \sim \mathrm{km s}^{-1}$ towards $\ell=271.9 \pm 2^0, b=29.6 \pm 1.4^0$, as is inferred from the observed dipolar modulation of the CMB temperature \citep{Kogut:1993ag,Akrami:2018vks}. Moreover diverse observations e.g.~\citet{lauer94,Hudson:2004et,Watkins:2008hf,Lavaux:2008th,Feldman:2009es,Colin:2010ds,Feindt:2013pma,6dFGSv16}, reaching out as far as $\sim300$~Mpc, have \emph{not} seen the expected $\sim 1/r$ fall-off of the peculiar velocity in the standard $\Lambda$CDM cosmology. The odds of this happening by chance in that framework can be estimated by querying Hubble volume simulations of large-scale structure formation e.g. Dark Sky~\citep{Skillman:2014qca}. In fact less than 1\% of Milky Way-like observers should observe the bulk flow ($>250~\mathrm{km~s}^{-1}$ extending to $z > 0.03$) that we do~\citep{Rameez:2017euv}. Thus we are \emph{not} comoving observers but are `tilted' relative to the idealised Hubble flow~\citep{Tsagas:2009nh}. The implications of this have been discussed for measurements of the Hubble parameter $H_0$~\citep{Hess:2014yka}, but not for the inference of cosmic acceleration.

Since cosmological observables are formulated in the `CMB frame' in which the Universe is supposedly perfectly isotropic, it is in any case always necessary to correct what we measure from our relative moving frame. For example the observed redshifts of the Type Ia supernovae (SNe~Ia) in catalogues like JLA~\citep{Betoule:2014frx} have been corrected in order to convert from the heliocentric frame to the CMB frame. The methodology used follows earlier work~\citep{Conley:2011ku} which used bulk flow observations made back in 2004~\citep{Hudson:2004et} and moreover assumed that there is convergence to the CMB frame beyond 150~Mpc. Since this is not in accordance with subsequent deeper observations, we use only the heliocentric redshifts and \emph{reverse} the corrections applied to the magnitudes in order to examine whether the deceleration parameter measured in our (bulk flow) rest frame can indeed differ from that of comoving observers in the model universe~\citep{Tsagas:2009nh}. Such theoretical considerations imply~\citep{Tsagas:2011wq,Tsagas:2015mua} that there should be a dipole asymmetry in the derived cosmic deceleration parameter $q_0$ towards the bulk flow direction. In this work we do find a significant ($3.9\sigma$) indication of such a dipole, and also that the monopole in $q_0$ decreases \emph{simultaneously} in significance (to $1.4\sigma$). Hence not only is the indication for acceleration statistically marginal~\citep{Nielsen:2015pga}, it probably arises due to our being tilted observers located in a bulk flow, rather than being the effect of a cosmological constant or dark energy.

\section{The `Joint Lightcurve Analysis'}
\label{sec:JLA}

We use the most up to date \emph{publicly} available sample of supernova lightcurve properties and directions: the SDSS-II/SNLS3 `Joint Lightcurve Analysis' (JLA) catalogue~\citep{Betoule:2014frx}. This consists of 740 spectroscopically confirmed SNe~Ia, including several low redshift ($z<0.1$) samples, three seasons of SDSS-II ($0.05 < z < 0.4$) and three years of SNLS ($0.2<z<1)$ data, all calibrated consistently in the `Spectral Adaptive Lightcurve Template 2' (SALT2) scheme. This assigns to each supernova 3 parameters: the apparent magnitude $m^*_B$ at maximum 
in the rest frame `$B$-band', and the light curve shape and colour corrections, $x_1$ and $c$. The distance modulus is then given by:
\begin{equation}
    \label{eq:DMOD}   
\mu_\mathrm{SN} = m^*_B - M + \alpha x_1 - \beta c ,
\end{equation}
where $\alpha$ and $\beta$ are assumed to be constants, as is $M$ the absolute SNe~Ia magnitude, as befits a `standard candle'. In the
standard $\Lambda$CDM cosmological model this is related to the luminosity distance $d_\textrm{L}$ as:
\begin{eqnarray}
& \mu &\equiv 25 + 5 \log_{10}(d_\textrm{L}/\textrm{Mpc}), 
 \quad \textrm{where:} \nonumber \\
& d_\textrm{L} &= (1 + z) \frac{d_\textrm{H}}{\sqrt{\Omega_k}} 
 \textrm{sin}\left(\sqrt{\Omega_k} \int_0^{z} \frac{H_0 \textrm{d}z'}{H(z')}\right), \textrm{for } \Omega_k > 0
 \nonumber \\
&  &= (1 + z) d_\textrm{H} \int_0^{z} \frac{H_0 \textrm{d}z'}{H(z')}, \textrm{for } \Omega_k = 0
 \nonumber \\
&  &= (1 + z) \frac{d_\textrm{H}}{\sqrt{\Omega_k}} 
 \textrm{sinh}\left(\sqrt{\Omega_k} \int_0^{z} \frac{H_0 \textrm{d}z'}{H(z')}\right), \textrm{for } \Omega_k < 0
 \nonumber \\
& d_\textrm{H} &= c/H_0, \quad H_0 \equiv 
 100h~\textrm{km}\,\textrm{s}^{-1}\textrm{Mpc}^{-1}, \nonumber \\
& H &= H_0 \sqrt{\Omega_\textrm{M} (1 + z)^3 + \Omega_k (1 + z)^2 
 + \Omega_\Lambda}.
\label{DLEQ}
\end{eqnarray}
Here $d_\textrm{H}$ is the Hubble distance and $H$ the Hubble parameter ($H_0$ being its present value), and $\Omega_\textrm{M}, \Omega_\Lambda, \Omega_k$ are the matter, cosmological constant and curvature densities in units of the critical density. In the $\Lambda$CDM model these are related by the `cosmic sum rule': $1=\Omega_\textrm{M} + \Omega_\Lambda + \Omega_k$. However we make no such model assumptions and will simply expand the luminosity distance $d_\textrm{L}$ in a Taylor series in order to examine its second derivative i.e. the acceleration (see \S \ref{sec:analysis}). This is because acceleration is a \emph{kinematic} quantity and can be measured without making any assumptions about the dynamics underlying the universal expansion. (There may be concern that such a Taylor expansion fails at high redshift, however we have verified that $d_\textrm{L}$ in the best-fit $\Lambda$CDM model differs by only 7\% even at $z=1.3$ (the highest redshift in the JLA sample), which is much less than the measurement uncertainty. Indeed the Taylor expansion fits the data just as well as $\Lambda$CDM.)

\begin{figure}
  \includegraphics[width=\columnwidth]{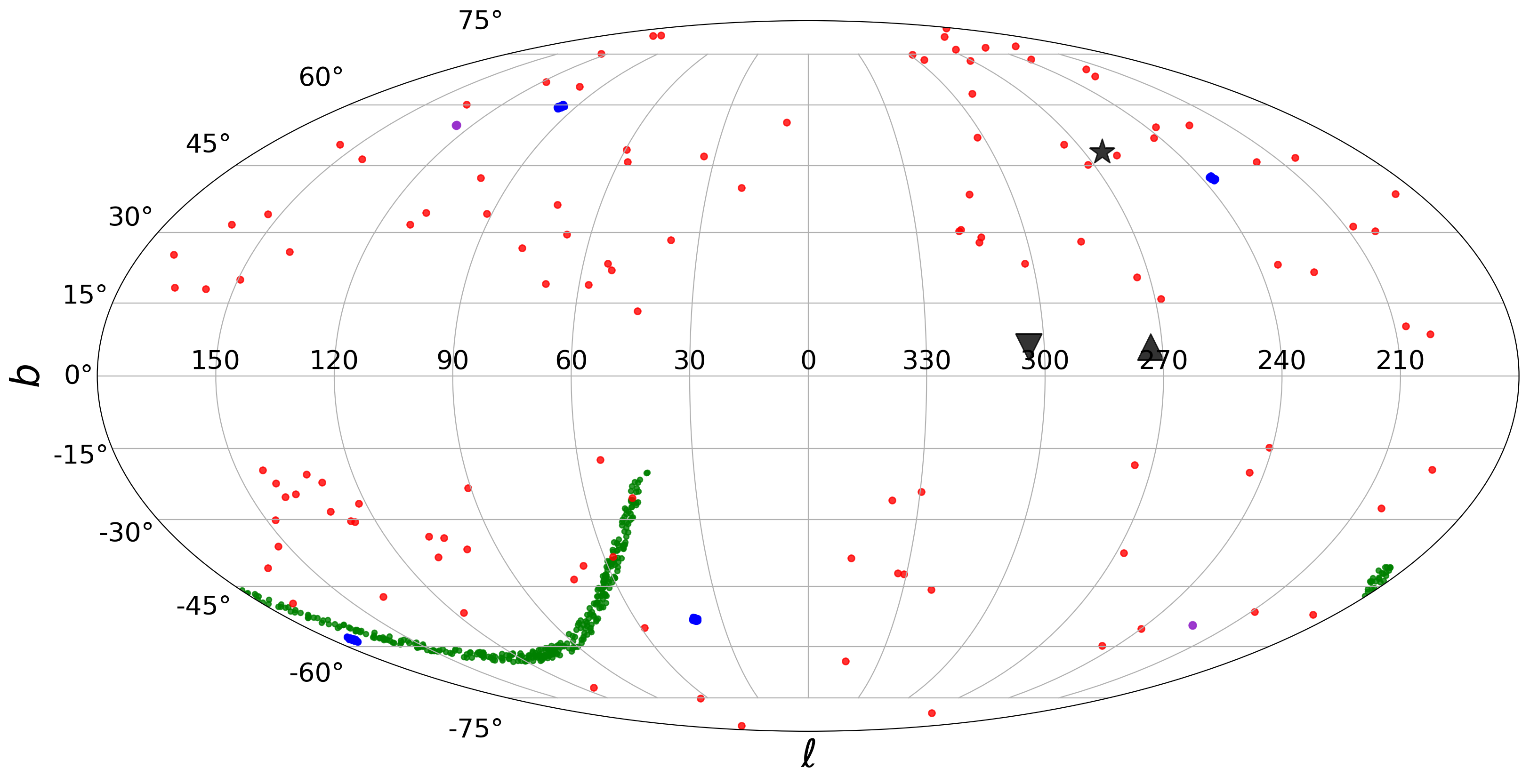}
  \caption{\small The sky distribution of the 4 sub-samples of the JLA catalogue in Galactic coordinates: SDSS (red dots), SNLS (blue dots), low redshift (green dots) and HST (black dots). Note that the 4 big blue dots are clusters of many individual SNe~Ia. The directions of the CMB dipole (star), the SMAC bulk flow (triangle) and the 2M++ bulk flow (inverted triangle) are also shown.}
  \label{fig:JLASkyScatter}
\end{figure}

Figure~\ref{fig:JLASkyScatter} is a Mollewide projection of the directions of the 740 SNe~Ia in Galactic coordinates. Due to the diverse survey strategies of the sub-samples that make up the JLA catalogue, its sky coverage is patchy and anisotropic. While the low redshift objects are spread out unevenly across the sky, the intermediate redshift ones from SDSS are mainly confined to a narrow disk at low declination, while the high redshift ones from SNLS are clustered along 4 specific directions. 

The JLA analysis~\citep{Betoule:2014frx} corrects the observed redshifts in the heliocentric frame, $z_\mathrm{hel}$, in order to obtain the cosmological redshifts, $z_\mathrm{CMB}$, after accounting for peculiar motions in the local Universe. These corrections are carried over unchanged from an earlier analysis~\citep{Conley:2011ku}, which in turn cites an earlier method~\citep{Neill:2007fh} and the peculiar velocity model of Hudson \emph{et al.}~\citep{Hudson:2004et}. It is stated that the inclusion of these corrections allow SNe~Ia with redshifts down to 0.01 to be included in the cosmological analysis, in contrast to earlier analyses~\citep{Riess:2006fw} which employed only SNe~Ia down to $z=0.023$.

\begin{figure}
 \center{\includegraphics[width=0.75\columnwidth]{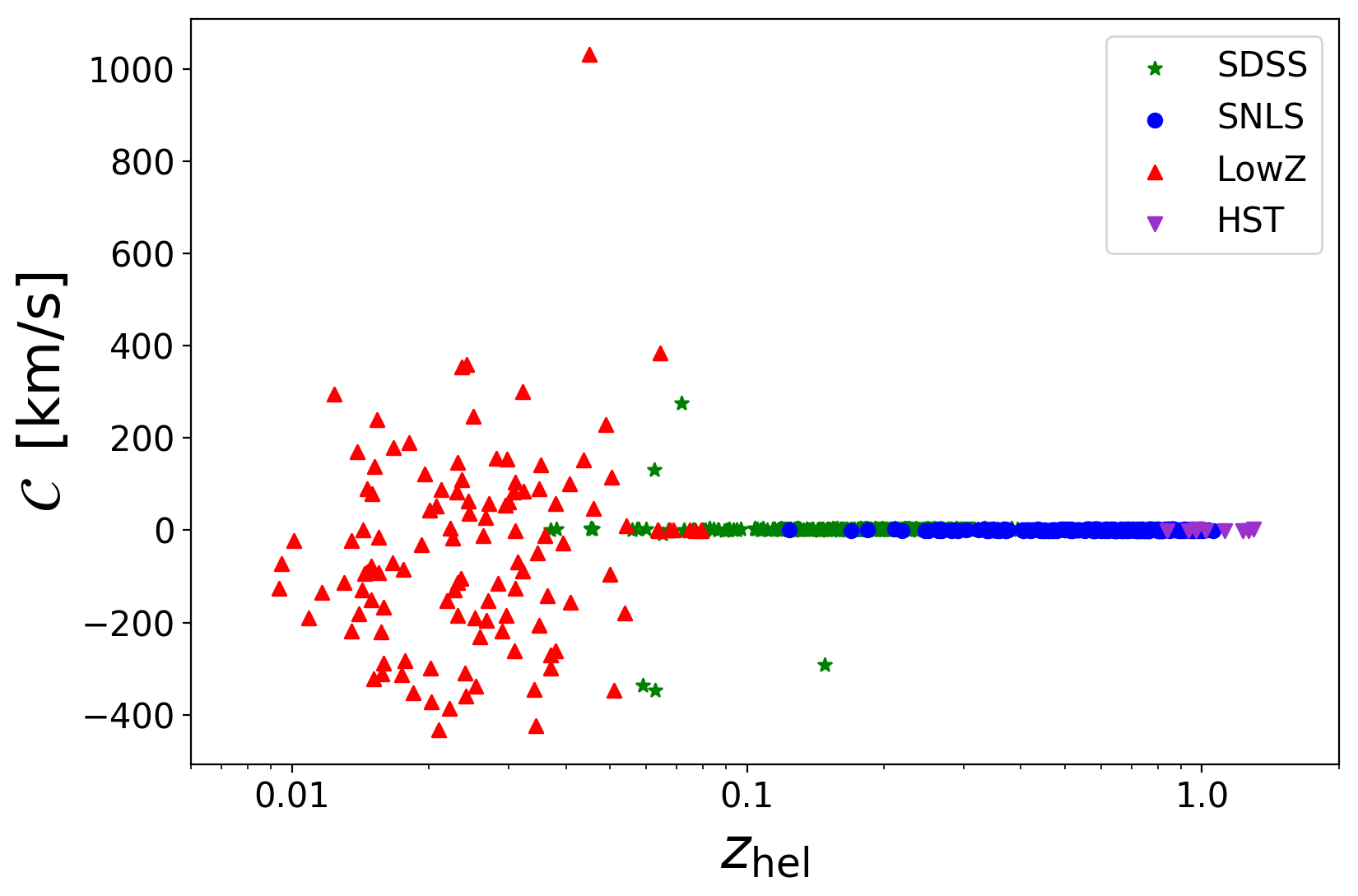}}
 \caption{\small The peculiar velocity `corrections' which have been applied to the JLA catalogue. The velocity parameter ${\cal C}$  defined in equation~(\ref{eq:t3def}) is shown versus the observed redshift $z_\textrm{hel}$ in the heliocentric frame.} 
 \label{fig:JLAVelLos}
\end{figure}

In Figure~\ref{fig:JLAVelLos} we scrutinise these corrections by exhibiting the velocity parameter ${\cal C}$, defined as
\begin{equation}
    \label{eq:t3def}
    {\cal C} = [(1+z_\mathrm{hel}) - (1 + z_\mathrm{CMB})(1+z_\mathrm{d})] \times c
\end{equation}
where $z_\mathrm{hel}$ and $z_\mathrm{CMB}$ are as tabulated by JLA, while $z_\mathrm{d}$ is given by~\citep{Davis:2010jq}
\begin{equation}
    \label{eq:zdipdef}
    z_\mathrm{d} = \sqrt{\frac{1 - \vec{v}_{\mathrm{CMB} - \odot}.\hat{n}/c}{1 + \vec{v}_{\mathrm{CMB} - \odot}.\hat{n}/c }} -1,
\end{equation}
where $\vec{v}_{\mathrm{CMB} - \odot}$ is 369~km~s$^{-1}$ in the direction of the CMB dipole,\citep{Kogut:1993ag} and $\hat{n}$ is the unit vector in the direction of the supernova. It can be seen in Figure~\ref{fig:JLAVelLos} that SNe~Ia beyond $z\sim0.06$ have been \emph{assumed} to be stationary w.r.t. the CMB rest frame, and corrections applied only to those at lower redshifts. It is not clear how these corrections were made beyond $z \sim 0.04$, which is the maximum extent to which the Streaming Motions of Abell Clusters (SMAC) sample~\citep{Hudson:2004et} extends. This has a bulk velocity of $687 \pm 203$~km~s$^{-1}$  towards $\ell=260 \pm 13^0, b = 0 \pm 11^0$ out to $z=0.04$ at 90\%C.L., and a bulk velocity of $372 \pm 127$~km~s$^{-1}$  towards $\ell=273^0, b = 6^0$ generated by sources beyond 200~$h^{-1}$ Mpc ($\Rightarrow z \simeq 0.064$) at 98\% C.L. If the peculiar velocity field is not discontinuous, the SNe~Ia immediately outside this volume should have comparable velocities. Figure~\ref{fig:JLAVelLos} indicates however that the JLA peculiar velocity `corrections' have arbitrarily \emph{assumed} that the bulk flow abruptly disappears at this point! The JLA analysis~\citep{Betoule:2014frx} allows SNe~Ia beyond this distance to only have an uncorrelated velocity dispersion of $c\sigma_z = 150$~km~s$^{-1}$. In the absence of any evidence of convergence to the CMB rest frame, this assumption is unjustified since it is it quite possible that the observed bulk flow stretches out to much larger scales. In fact there have been persistent claims of a `dark flow' extending out to several hundreds of Mpc \citep{Kashlinsky:2008ut,Kashlinsky:2009dw,Kashlinsky:2010ur}, although this is still under debate. At any rate the value of $c\sigma_z$ should be determined by fitting to the data, rather than put in by hand.

At this point it is worth noting the anisotropy of the JLA catalogue. Out of the 740 SNe~Ia, 551 are in the hemisphere pointing \emph{away} from the CMB dipole. With respect to the $372 \pm 127$~km~s$^{-1}$ bulk flow of the model~\citep{Hudson:2004et} using which the redshifts of the local SNe~Ia have been corrected, only 108 are in the upper hemisphere while 632 are in the lower hemisphere. 
With respect to the direction of the abnormally high flow reported by 6dFGSv, the largest and most homogeneous peculiar velocity sample of nearly 9000 galaxies~\citep{6dFGSv16}, 103 SNe~Ia are in the upper hemisphere while 637 are in the lower hemisphere.

The subsequent Pantheon catalogue\citep{Scolnic:2017caz}, which incorporates 308 additional SNe~Ia (many from the Pan-STARRS survey), continues to suffer from these problems. While the flow model \citep{Carrick:2015xza} using which the redshifts of the Pantheon sample have been corrected go out to $z\sim0.067$, this model has a residual bulk flow of $159\pm 23$~km~s$^{-1}$ attributed to sources beyond $z = 0.067$, and 890 of the 1048 Pantheon SNe are in the hemisphere \emph{opposite} to the direction of this flow. 

Both JLA and Pantheon include SNe to which anomalously large peculiar velocity corrections have been applied at redshifts far higher than the limit to which the corresponding flow models extend. Two of the many such examples are SDSS2308 in JLA at $z=0.14$ in JLA (the outlier in Figure~2) and SN2246 in Pantheon at $z=0.194$.

We use the heliocentric redshifts tabulated by JLA~\citep{Betoule:2014frx} and subtract out the bias corrections applied to $m^*_B$. For the Pantheon catalogue~\citep{Scolnic:2017caz} the $z_\mathrm{hel}$ values and individual contributions to the covariance are not public, and moreover there are unresolved  concerns about the accuracy of the data therein~\citep{Rameez:2019nrd} so we cannot use it.

\section{Cosmological analysis}
\label{sec:analysis}

We now compare the distance modulus (equation~\ref{eq:DMOD}) obtained from the JLA sample with the apparent magnitude (equation~\ref{DLEQ}) using the Maximum Likelihood Estimator~\citep{Nielsen:2015pga}. For the luminosity distance we use its kinematic Taylor series expansion up to the third term~\citep{Visser:2003vq} since we wish to analyse the data without making assumptions about the matter content or the dynamics:

\begin{equation}
\label{eq:Qkin}    
d_\textrm{L} (z_\mathrm{hel}) = \frac{cz_\mathrm{hel}}{H_0} \bigg\{ 1 + \frac{1}{2}[1 - q_0]z_\mathrm{hel} - \frac{1}{6} [1-q_0 - 3q^2_0 + j_0 + \frac{kc^2}{H^2_0 a^2_0} ] z_\mathrm{hel}^2 \bigg\}
\end{equation}
where $q \equiv -\ddot{a} a/\dot{a}^2$ is the cosmic deceleration parameter in the Hubble flow frame, defined in terms of the scale factor of the universe $a$ and its derivatives w.r.t. proper time, $j_0$ is the cosmic `jerk' $j = \dot{\ddot{a}}/aH^3$, and $-kc^2/(H^2_0 a^2_0)$ is just $\Omega_k$. Note that  the last two appear together in the coefficient of the $z^3$ term so cannot be determined separately. In the $\Lambda$CDM model: $q_0 \equiv \Omega_\mathrm{M}/2 - \Omega_\Lambda$.

To look for a dipole in the deceleration parameter, we allow it to have a direction dependence:
\begin{equation}
\label{eq:qdef2}
    q = q_\mathrm{m} + \vec{q}_\mathrm{d}.\hat{n}\mathcal{F}(z, S)
\end{equation}
where $q_\mathrm{m}$ and $q_\mathrm{d}$ are the monopole and dipole components, while $\hat{n}$ is the direction of the dipole and $\mathcal{F}(z, S)$ describes its scale dependence. We consider four representative functional forms:\\ 
(a) Constant: $\mathcal{F}(z, S) =1$ independent of $z$,\\ 
(b) Top hat: $\mathcal{F}(z, S) = 1$ for $z<S$, and $0$ otherwise,\\ 
(c) Linear: $\mathcal{F}(z,S) = 1-z/S$, and\\ 
(d) Exponential: $\mathcal{F}(z,S) = \exp(-z/S)$.\\ 
Due to the anisotropic sky coverage of the dataset, it would be hard to find $\hat{n}$ from the data, so we choose it to be along the CMB dipole direction. This is reasonable as the directions of the reported bulk flows~\citep{Hudson:2004et,Watkins:2008hf,Lavaux:2008th,Feldman:2009es,Colin:2010ds,Feindt:2013pma,6dFGSv16} are all within $\sim 40^0$ of each other and of the CMB dipole. (Later we allow the direction to vary as an \emph{a posteriori} test to demonstrate that our result is indeed robust.)

We maximise a likelihood constructed earlier~\citep{March:2011xa,Nielsen:2015pga}, simultaneously with respect to the 4 cosmological parameters $q_\mathrm{m}, j_0-\Omega_k, q_\mathrm{d}$ and $S$, as well as the 8 parameters that go into the standardisation of the SNe~Ia candles: $\alpha, \beta, M_0, \sigma_{M_0}, x_{1,0}, \sigma_{x_{1, 0}}, c_0$ and $\sigma_{c_0}$. While our analysis is frequentist, it is equivalent to the `Bayesian Hierarchical Model' of March \emph{et al.}~\citep{March:2011xa} (which indeed yielded the same result~\citep{Shariff:2015yoa} as the frequentist analysis of \citet{Nielsen:2015pga} when applied to the JLA catalogue).

\begin{table*}
\tiny{
\caption{The tilted local universe, with $\sigma_z$ set to zero, fitted to data with the MLE.} 
\centering
\begin{tabular} 
{ l  c  c  c  c  c  c  c  c  c  c  c  c  c }
 \hline\hline
 & -2 log ${\cal L}_\textrm{max}$ & $q_\mathrm{m}$ & $q_\mathrm{d}$ & $S$ & $j_0-\Omega_k$ & $\alpha$ & $x_{1,0}$ & $\sigma_{x_{1,0}}$ & $\beta$ & $c_0$ & $\sigma_{c_0}$ & $M_0$ & $\sigma_{M_0}$ \\ 
\hline
Tilted universe & -208.28 & -0.157 & -8.03 & 0.0262 & -0.489 & 0.135 & 0.0394 & 0.931 & 3.00 & -0.0155 & 0.071 & -19.027 & 0.114\\  
No tilt ($q_\mathrm{d}=0$) & -189.52 & -0.166 & 0 & - & -0.460& 0.133 & 0.0396 & 0.931 & 2.99 & -0.014 & 0.071 & -19.028 & 0.117\\
No accn. ($q_\mathrm{m}= 0$) & -205.98 & 0  & -6.84 & 0.0384 & -0.836 & 0.134 & 0.0365 & 0.931 & 2.99 & -0.014 & 0.071 & -19.002 & 0.115\\
 \hline 
\end{tabular}
\tablefoot{The BIC for the models above is -129.00, -123.45 and -133.31, providing `strong' evidence for the last model.}
\label{tab:results3}
}
\end{table*}

\begin{table*}
\tiny{
\caption{The tilted local universe (with $\sigma_z$ set to zero) fitted to data with the ``constrained $\chi^2$ method''.} 
\centering
\begin{tabular} 
{ l  c  c  c  c  c  c  c  c  c  c  c  c  c }
 \hline\hline
 & $q_\mathrm{m}$ & $q_\mathrm{d}$ & $S$ & $j_0-\Omega_k$ & $\alpha$ & $\beta$ & $M_0$ & $\sigma_{\rm int}$ \\ 
\hline
Tilted universe &  -0.268 & -6.54 & 0.0297 & -0.517 & 0.135 &  3.04 &  -19.053 & 0.124\\  
No tilt ($q_\mathrm{d}=0$) & -0.307 & 0 & - & -0.523 & 0.133 & 3.03 & -19.047 & 0.133\\
 \hline 
\end{tabular}
\label{tab:resultsCC}
}
\end{table*}

\begin{table*}
\tiny{
\caption{The tilted local universe, with $\sigma_z$ left floating, fitted to data with the MLE.}
\centering
\begin{tabular} 
{ l  c  c  c  c  c  c  c  c  c  c  c  c  c  c }
 \hline\hline
 &-2 log ${\cal L}_\textrm{max}$ & $q_\mathrm{m}$ & $q_\mathrm{d}$ & $S$ &  $j_0-\Omega_k$ & $\alpha$ & $x_{1,0}$ & $\sigma_{x_{1,0}}$ & $\beta$ & $c_0$ & $\sigma_{c_0}$ & $M_0$ & $\sigma_{M_0}$ & $c\sigma_z$ [km/s]\\ 
\hline
Tilted universe& -216.90 & -0.154  & -6.33 & 0.0305 & -0.497 & 0.134 & 0.0395 & 0.932 & 3.04 & -0.0158 & 0.071 & -19.022 & 0.106 & 241\\  
No tilt ($q_\mathrm{d}=0$) & -203.23 & -0.187 & 0 & - & -0.425 & 0.133 & 0.0398 & 0.932 & 3.05 & -0.0151 & 0.071 & -19.032 & 0.106 & 274\\  
No accn. ($q_\mathrm{m} = 0$) & -214.74 & 0 & -5.60 & 0.0350 & -0.833 & 0.133 & 0.0368 & 0.932 & 3.04 & -0.0145 & 0.071 & -19.000 & 0.106 & 243\\  
\hline
\end{tabular}
\tablefoot{The BIC for the models above is -131.01, -130.55 and -135.46, providing `positive' evidence for the last model.}
\label{tab:results4}
}
\end{table*}

In Table~\ref{tab:resultsCC} we show how this compares with using the prevalent ``constrained $\chi^2$ method'' used by e.g. \cite{Betoule:2014frx}, wherein an arbitrary error $\sigma_{\rm int}$ is added to each data point and varied until a good fit (with $\chi^2=1$/d.o.f.) is obtained to the \emph{assumed} model. This may be appropriate for parameter estimation, but \emph{not} for model selection.

However as seen in Table~\ref{tab:results3} the quality of fit \emph{improves} further (-2 log $\mathcal{L}_\mathrm{max}$ decreases) when $q_0$ is allowed to have a dipole. In the best fit where this has an exponentially decaying form $\propto \mathrm{e}^{-z/S}$, the dipole $q_\mathrm{d}=-8.03$ is much \emph{larger} than the monopole $q_\mathrm{m}=-0.157$ and its scale parameter is $S = 0.0262$ indicating that the impact of the bulk flow dominates over any isotropic acceleration out to $z \sim0.1$. Since $\Delta_\textrm{BIC}$ between the model with $q_\mathrm{d} = 0$ and the model with $q_\mathrm{m} = 0$ is 9.86, this constitutes `strong' evidence against a universe that is accelerating isotropically. In the presence of this dipole, $q_\mathrm{m} = 0$ is disfavoured at only $1.4\sigma$. In other words, in a universe where we have theoretical reasons to expect a dipolar modulation in the deceleration parameter in the direction of our motion through the CMB, there is \emph{no} significant evidence for a non-zero value of its monopole component. Figure~4 shows the 1, 2 and 3 $\sigma$ contours in the likelihood around the maximum as a function of $q_\mathrm{d}$ and $q_\mathrm{m}$, profiling over all other parameters. 

\begin{figure}
 \center{\includegraphics[width=0.75\columnwidth]{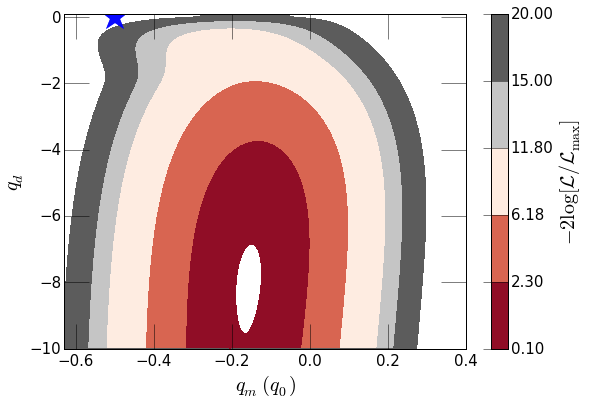}}
 \caption{\small 1, 2 and 3 $\sigma$ contours (corresponding to -2 log ${\cal L} / {\cal L}_\mathrm{max}$ = 2.3, 6.18 and 11.8 respectively) for the monopole and dipole components of the cosmological deceleration parameter inferred from the JLA catalogue of 740 SNe~Ia, profiling over all other parameters. Note that the vertical scale for the magnitude of the dipole is compressed by x10 relative to the horizontal scale for the monopole. The value of $q_0$ for the standard $\Lambda CDM$ model is shown as a blue star.}
 \label{fig:QExpContourScan}
\end{figure}

We also study the effect of allowing an additional uncorrelated velocity dispersion $c\sigma_z$ in the fit --- rather than fixing it to be $150$~km~s$^{-1}$ as in the JLA analysis~\citep{Betoule:2014frx}. As shown in Table~\ref{tab:results4} this improves the overall fit even further for $c\sigma_z=241$~km~s$^{-1}$; the best-fit dipole drops a little to $q_\mathrm{d}=-6.33$, while the monopole is nearly unchanged at $q_\mathrm{m}=-0.154$. The $\Delta_\textrm{BIC}$ between the model with  $q_\mathrm{d} = 0$ and the one with $q_\mathrm{m} = 0$ is 4.91, providing `positive' evidence against a universe that is accelerating isotropically. Our main result is thus robust in that the maximum likelihood estimator prefers to interpret the data as evidence of a dipole in the deceleration parameter aligned with the CMB dipole, rather than as an isotropic acceleration of the universe which may indicate the presence of a cosmological constant.

\begin{figure}
 \center{\includegraphics[width=0.45\columnwidth]{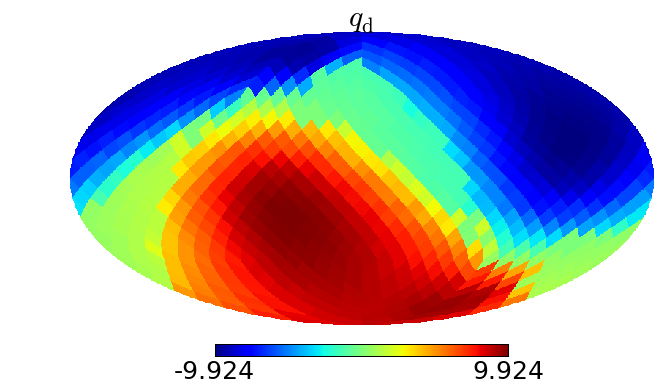}\includegraphics[width=0.45\columnwidth]{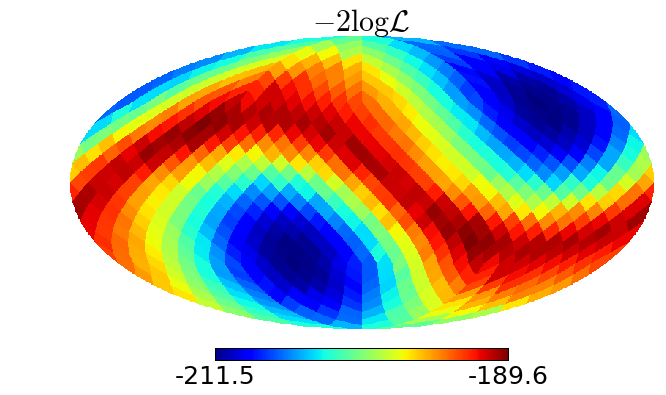}}
 \caption{Results of an \emph{a posteriori} grid scan (left panel) varying the direction of the scale-dependent dipolar modulation of the form $q_0 = q_\mathrm{m} + \vec{q}_\mathrm{d}.\hat{n}\exp(-z/S)$ in Galactic coordinates. The best-fit direction is within $23^0$ of the CMB dipole and -2 log ${\cal L}$ (right panel) changes by just 3.22 between these two directions.}
 \label{fig:LLHandQdipDirscan}
\end{figure}

As an \emph{a posteriori} test, we examine the direction dependence of this scale-dependent dipolar modulation in $q_0$, by scanning the direction of $q_d$ on a grid corresponding to a HEALpix~\citep{Gorski:2004by} map of nside=8. The best fit direction is 23 degrees away from the CMB dipole, where  $q_\mathrm{d}$ increases to -9.851 but -2 log ${\cal L}$ improves by only 3.22. This demonstrates that the direction of the anisotropy  we find is also robust.

\section{Discussion}

It has been observed~\citep{Bernal:2017} that the deceleration parameter inferred from previous SNe~Ia datasets has a redshift, and indeed directional, dependence. This was interpreted as indicative of local anisotropy in the matter distribution, i.e. our being located in an asymmetric void. The refinement in the present work is that we consider a recent comprehensive database of SNe~Ia and take into account all systematic effects as encoded in the covariance matrices provided~\citep{Betoule:2014frx}. Moreover we focus on the local \emph{velocity} rather than the density field as this fully reflects the gravitational dynamics due to inhomogeneities. We can then explore the expected consequences of our being `tilted' i.e. \emph{non}-Copernican observers. Our analysis is guided by the suggestion that we may then infer acceleration even when the overall expansion rate is  decelerating --- a signature of which would be a dipolar modulation of the inferred $q_0$ along the direction of the bulk flow~\citep{Tsagas:2009nh,Tsagas:2011wq,Tsagas:2015mua}.

The effect of peculiar velocities on SNe~Ia cosmology has been discussed earlier~\citep{Hui:2005nm,Davis:2010jq}, however these studies relied on covariances that apply to Copernican observers. As we show elsewhere~\citep{Colin:2019}, the bulk flow we are embedded in is rare at a level of $\lesssim1\%$~\citep{Rameez:2017euv} according to the DarkSky simulation~\citep{Skillman:2014qca}, but the conditional covariances can be up to a factor of $\sim10$ larger and introduce a preferred direction locally. This can make a much bigger impact on cosmological inferences than was found in previous studies. In particular the JLA analysis~\citep{Betoule:2014frx} of the \emph{same} dataset claimed that the effect of peculiar velocities is a tiny ($<0.1\%$) shift in the best-fit cosmological parameters.

In summary, the model-independent evidence for acceleration of the Hubble expansion rate from the largest public catalogue of Type~Ia supernovae is only $1.4\sigma$. This is in contrast to the claim~\citep{Scolnic:2017caz} that acceleration is established by SNe~Ia at $>6\sigma$ in the framework of the $\Lambda$CDM model. Moreover there is a significant ($3.9\sigma$) indication for a dipole in $q_0$ towards the CMB dipole --- as is indeed expected if the apparent acceleration is an artefact of our being located in a local bulk flow which extends out far enough to include most  of the supernovae studied~\citep{Tsagas:2009nh,Tsagas:2011wq,Tsagas:2015mua}. Given the observational evidence that there is \emph{no} convergence to the CMB frame as far out as redshift $z \sim 0.1$ which includes half the known SNe~Ia, this possibility must be taken seriously. 


\begin{acknowledgements}
We thank the JLA collaboration for making \emph{all} their data public and Dan Scolnic for correspondence concerning the Pantheon catalogue. We are grateful to Mike Hudson and Christos Tsagas for discussions. MR acknowledges a Carlsberg distinguished postdoctoral fellowship and hospitality at the Institut d'Astrophysique, Paris.
\end{acknowledgements}

\noindent
{\bf Code availability:} The code used here is available at: https://github.com/rameez3333/Dipole\_JLA.



\begin{appendix}

\section{Redshift-dependence of light curve fitting parameters}

\begin{table*}[h]
\tiny{
\caption{Fits to the JLA catalogue allowing for sample- and redshift-dependence of SNe~Ia parameters.}
\centering
\begin{tabular}{ l  c  c  c  c  c  c  c  c  c  c  c  c  c  c  c }
\hline\hline
& -2 log $\mathcal{L}_\mathrm{max}$ & $q_\mathrm{m}$  & $q_\mathrm{d}$ & $S$ & $j_0 - \Omega_k$ &$\alpha$ & $\beta$ & $M_0$ & $\sigma_{M_0}$\\
\hline
R \& H (22 parameters) with peculiar \\ velocity corrections and no dipole ($q_\mathrm{d}=0$) & -331.6 & -0.457 & 0 & -- & 0.146 & 0.135 & 3.07 & -19.074 & 0.107\\
-- as above, with no acceleration ($q_\mathrm{m}=0$) & -315.6 & 0 & 0 & -- & -1.35 & 0.132 & 3.05 & -19.013 & 0.109\\
\hline
R \& H (22 parameters) with no peculiar \\ velocity corrections and no dipole ($q_\mathrm{d}=0$) & -306.70 & -0.333 & 0 & -- & -0.397 & 0.133 & 3.00 & -19.050 & 0.116\\
-- as above, with no acceleration ($q_\mathrm{m}=0$) & -298.15 & 0 & 0 & -- & -1.37 & 0.132 & 2.98 & -19.011 & 0.117\\
\hline
R \& H (24 parameters) with no peculiar \\ velocity corrections and dipole $\propto \textrm{e}^{-z/ S}$ & -325.00 & -0.310 & -8.09 & 0.0256 & -0.471 & 0.135 & 3.01 & -19.053 & 0.113\\
-- as above with no dipole ($q_\mathrm{d}=0$) & -306.70 & -0.333 & 0 & 0 & -0.400 & 0.134 & 3.00 & -19.054 & 0.116\\
-- as above, with no acceleration ($q_\mathrm{m}=0$) & -318.14 & 0 & -6.19 & 0.0344 & -1.32 & 0.133 & 3.00 & -19.012 & 0.114\\
\hline
\end{tabular}
\tablefoot{The first row corresponds to the 22-parameter fit of \citet{Rubin:2016iqe} with the full JLA covariances. In the second row, the peculiar velocity corrections they made are undone (and the corresponding arbitrary uncertainties imposed are excluded). The third row demonstrates the dramatic improvement in the fit when a scale-dependent exponentially falling dipole (with 2 additional parameters) is allowed for in $q_0$.}
\label{tab:RandHreprodresults}
}
\end{table*}

In this work we have used the statistical approach as well as the treatment of lightcurve parameters espoused by \citet{Nielsen:2015pga}. These authors were criticised by \citet{Rubin:2016iqe} for using redshift-\emph{independent} distributions for $x_1$ and $c$. In this respect we note the following:

\begin{enumerate}

\item The JLA analysis determined the relationship between the luminosity distance and redshift for SNe~Ia. To inspect \emph{a posteriori} the distribution of two ($x_1$ and $c$) out of the three ingredients that go into standardising SNe~Ia, and then add empirical terms  in the fit to describe their sample dependence and redshift evolution, is fundamentally against the principles of blind hypothesis testing, especially since no such dependence had been suggested by~\citet{Betoule:2014frx}. 

\item Nevertheless we carry out the same 22-parameter fit~\citep{Rubin:2016iqe} for comparison and present the results in Table~\ref{tab:RandHreprodresults}. 
While the log maximum likelihood ratio does improve for these fits, this parameterisation actually \emph{increases} the significance of the dipole in $q_0$ to $4.7\sigma$ (likelihood ratio of 18.3) and reduces further the significance of a monopole.

\item The addition of parameters to improve the quality of a fit and obtain a desired outcome have to be justified by physical and/or information theoretic arguments. The additional parameters of \citet{Rubin:2016iqe} can be justified by the Akaike information criterion but not by the stricter Bayesian information criterion. This also applies to the two additional parameters we introduce ($q_\mathrm{d}$ and $S$) but these are physically motivated for a `tilted observer'~\citep{Tsagas:2011wq,Tsagas:2015mua}.

\item If the light curve parameters $x_1$ and $c$ are allowed to be sample/redshift dependent one can ask why the absolute magnitude of SNe~Ia should also not be sample/redshift-dependent. Allowing this of course undermines their use as `standard candles' and the data is then unsurprisingly consistent with \emph{no} acceleration~\citep{Tutusaus:2017ibk}, as seen in Table~\ref{tab:RandHreprodresults}. 

\end{enumerate}

\section{Uncertainties}
\label{sec:UNC}

\begin{figure}[h]
  \center{\includegraphics[width=0.75\columnwidth]{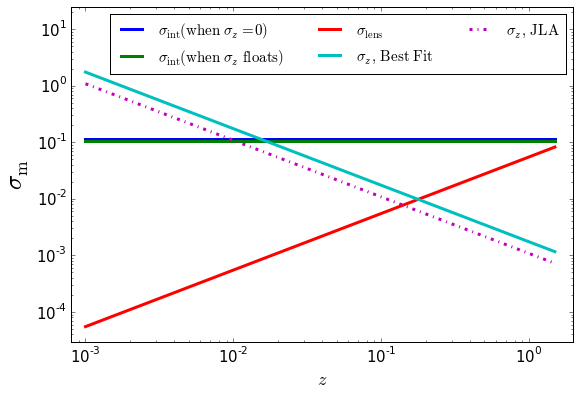}}
 \caption{\small The different sources of intrinsic dispersion in magnitude $\sigma_\mathrm{m}$ that enter cosmological fits of JLA data. On top of the global $\sigma_\mathrm{int}$, SNe~Ia are given a dispersion $\sigma_\mathrm{lens}$ proportional to redshift to account for lensing, while low redshift SNe~Ia are selectively more dispersed by $\sigma_z$ to account for  peculiar velocity effects.}
 \label{fig:DispBudget}
\end{figure}

The JLA covariance matrix includes uncertainties from the lightcurve template fitting process, calibration uncertainties, dust extinction in the Galaxy etc, together with the expected dispersion due to peculiar velocities (which mainly affects low redshift SNe) and lensing (which mainly affects high redshift SNe~Ia), as well as the propagated uncertainties from the flow model using which the SN by SN peculiar velocity corrections are performed. In addition it is also necessary to fit for a global intrinsic dispersion as in previous analyses~\citep{March:2011xa}.  We use heliocentric redshifts in this analysis so do not include uncertainties related to the peculiar velocity corrections based on the flow model. The redshift dependence of the dispersions in the fit are shown in Figure~\ref{fig:DispBudget}.

\end{appendix}

\end{document}